\documentclass[useAMS,usenatbib]{mn2e}
\usepackage{amssymb}
\usepackage{graphicx}
\usepackage{lscape}
\usepackage{rotating}
\begin{document}

\title[Spectra and pulse profiles of isolated neutron stars]{
The imprint of the crustal magnetic field on the thermal spectra and pulse profiles
of isolated neutron stars}

\author[R.~Perna et al.]{Rosalba Perna$^1$, Daniele Vigan\`o$^2$,
  Jos\'e A.~Pons$^2$, Nanda~Rea$^3$ \\ 
$^1$ Department of Astrophysical and Planetary Sciences and JILA,
  University of Colorado, 440 UCB, Boulder, 80309, USA\\ 
$^2$ Departament de F\'isica
  Aplicada, Universitat d'Alacant, Ap. Correus 99, 03080 Alacant,
  Spain\\ 
$^3$ Institute of Space Sciences (CSIC--IEEC), Campus UAB,
  Faculty of Science, Torre C5-parell, E-08193 Barcelona, Spain}

\date{}
\maketitle

\label{firstpage}

\begin{abstract}

Isolated neutron stars (NSs) show a bewildering variety of
astrophysical manifestations, presumably shaped by the magnetic field
strength and topology at birth.  Here, using state-of-the art
calculations of the coupled magnetic and thermal evolution of NSs, we
compute the thermal spectra and pulse profiles expected for a variety
of initial magnetic field configurations. In particular, we contrast
models with purely poloidal magnetic fields to models dominated by a
strong internal toroidal component. We find that, while the former
displays double peaked profiles and very low pulsed fractions, in the
latter, the anisotropy in the surface temperature produced by the
toroidal field often results in a single pulse profile, with pulsed
fractions that can exceed the $50-60\%$ level even for perfectly
isotropic local emission.  We further use our theoretical results to
generate simulated ``observed'' spectra, and show that blackbody (BB)
fits result in inferred radii that can be significantly smaller than
the actual NS radius, even as low as $\sim 1-2$~km for old NSs with
strong internal toroidal fields and a high absorption column density
along their line of sight. We compute the size of the inferred BB
radius for a few representative magnetic field configurations, NS
ages, and magnitudes of the column density.  Our theoretical results
are of direct relevance to the interpretation of X-ray observations of
isolated NSs, as well as to the constraints on the equation of state
of dense matter through radius measurements.

\end{abstract}

\section{Introduction}

Observations of isolated neutron stars (NSs) over the last several
decades have painted a zoo of disparate manifestations, in energy
bands ranging from the gamma-rays to the radio.  The bulk of 
isolated NSs manifest themselves as radio pulsars and are
characterized by a rather steady spin down.
These NSs have an estimated magnetic field in the $\sim 10^{12}-10^{13}$~G range.  A
fraction of NSs are further characterized by large surface
temperatures for their ages, occasional X-ray bursts, and, in some
cases, even giant $\gamma$-ray flares. These objects, historically
classified as Anomalous X-ray Pulsars (AXPs) and Soft $\gamma$-ray
Repeaters (SGRs), and collectively known as magnetars, have very large
estimated magnetic field strengths, $\sim 10^{14}-10^{15}$~G. The
large fields are believed to be responsible for their observational
characteristics \citep{thompson95}.  

Other NSs, at the opposite end of the spectrum, are very quiet and
their surface emission is generally consistent with thermal emission
from the entire surface of the star. These objects, also known as
Central Compact Objects (CCOs), have been proposed to be NSs with very
low external magnetic field strengths, $B\lesssim 10^{11}$~G, either by birth
\citep{gotthelf09}, or as the result of field screening by fallback
accreted matter \citep{muslimov95,young95,bernal10,ho11,vigano12b}.

Evidently, the magnetic field strength of a NS plays a fundamental
role in its observational appearance and in its evolutionary
path. Indeed, a large number of investigations over several decades
have been aimed at understanding how the $B$-field shapes the NS life
and properties as we see them { (e.g.~\citealt{heyl98})}.  Clearly,
the dipolar field component, inferred through measurements of $P$ and
$\dot{P}$, is not sufficient to account for this variety of
behaviours, calling for a re-evaluation of our global understanding of
the relation between the inferred $B$-field and the phenomenology of a
NS.  As discussed in a series of previous papers
\citep{pons07,aguilera08b,pons09,vigano13}, magnetic and thermal
evolution are strongly coupled.  The standard, 1D cooling theory is
able to predict the range of observed temperatures and luminosities
for weakly magnetised objects. However, under the presence of a strong
magnetic field, two other important effects play a role in influencing
the observational properties of the NSs: the gradual conversion of
magnetic energy into heat via the Joule effect in the crust, and the
anisotropy in the surface temperature distribution.

For strongly magnetised NSs ($B\gtrsim 10^{14}$ G), the dissipation of
currents circulating in the crust maintains the crust at high
temperatures ($T_c\gtrsim 10^8$ K) and consequently the average
surface temperature is also higher than in the non-magnetic case. The
first consequence is an enhancement of the persistent luminosity: the
most magnetized objects (magnetars, high-B pulsars) are systematically
brighter, and they are much easier to detect than the weakly
magnetised ones.  The second important effect, on which we focus in
this work, is the anisotropy in temperature induced by the magnetic
field. Under the presence of a strong field, the conductivity becomes
anisotropic, due to the fact that the electrons, which are the main
responsible for the heat transport, move more easily along the field
lines than across them. In the outer crust and in the envelope of
NSs, the magnetic field geometry drives the preferred
direction for the heat conduction, acting in some regions as a thermal
insulator.

In \cite{vigano13} (Paper I hereafter), we presented the most detailed 2D simulations 
of the magneto-thermal evolution of isolated NSs to date, including updated microphysics inputs.  
We showed that the observed timing
properties and bolometric luminosities of a varied sample of about 40
sources (including AXPs, SGRs, high-$B$ radio pulsars,
rotation-powered pulsars, X-ray isolated NSs, CCOs) can be accounted
for by varying the strength of the initial poloidal field, and, in
some cases, including the presence of a dominant toroidal field at
birth.

In this paper, we extend our study of the NS zoo by studying the
expected thermal spectra and pulse profiles predicted in a sample of
magnetothermal evolutionary scenarios. In particular, we will contrast
models with and without a dominant toroidal component of the magnetic
field. Our results demonstrate that
the thermal pulse profiles are markedly distinct in the two cases, hence
allowing a probe of the magnetic field geometry.

We further address the issue of the magnitude of the
blackbody radius and temperature inferred from fits to single blackbody
models, when the underlying temperature profile is anisotropic.

Our paper is organized as follows: in \S~\ref{sec:theory}, we
discuss the origin of the surface temperature anisotropy and show results
from a set of simulations and ensuing surface temperature profiles. In
\S~\ref{sec:spectra}, we use the latter to build synthetic spectra,
and compute the effective blackbody radius that would be inferred in
fits; we also compute pulse profiles, for the same theoretical
temperature profiles, in a soft X-ray thermal band. In \S~\ref{sec:obs}, we
discuss our results in the context of observations. We summarize and
conclude in \S~\ref{sec:conclusion}.

\section{Surface temperature profiles}\label{sec:theory}

\begin{figure}
\vspace{-0.1in}
\includegraphics[width=.4\textwidth]{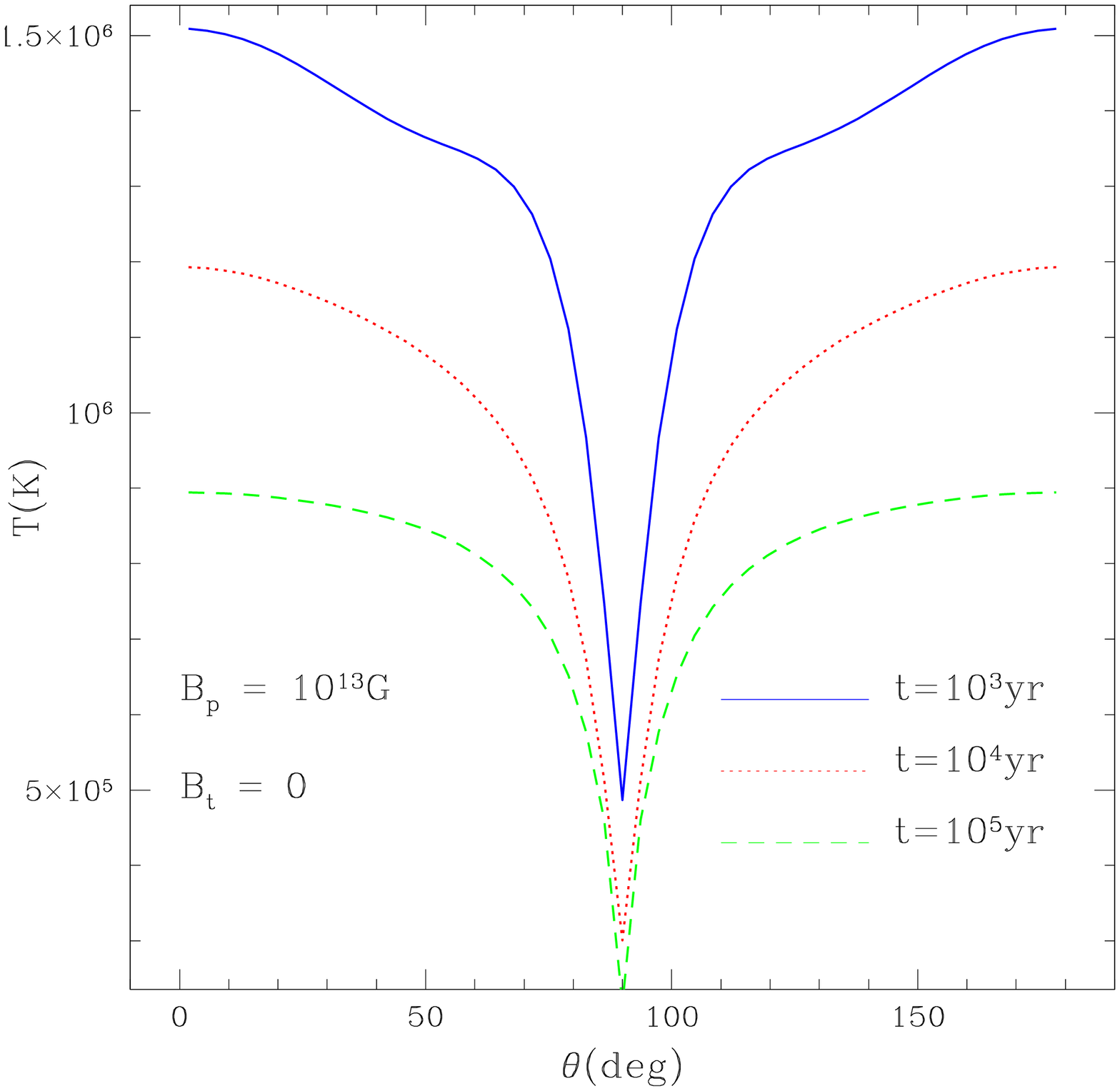}\\
\vspace{-0.1in}
\includegraphics[width=.4\textwidth]{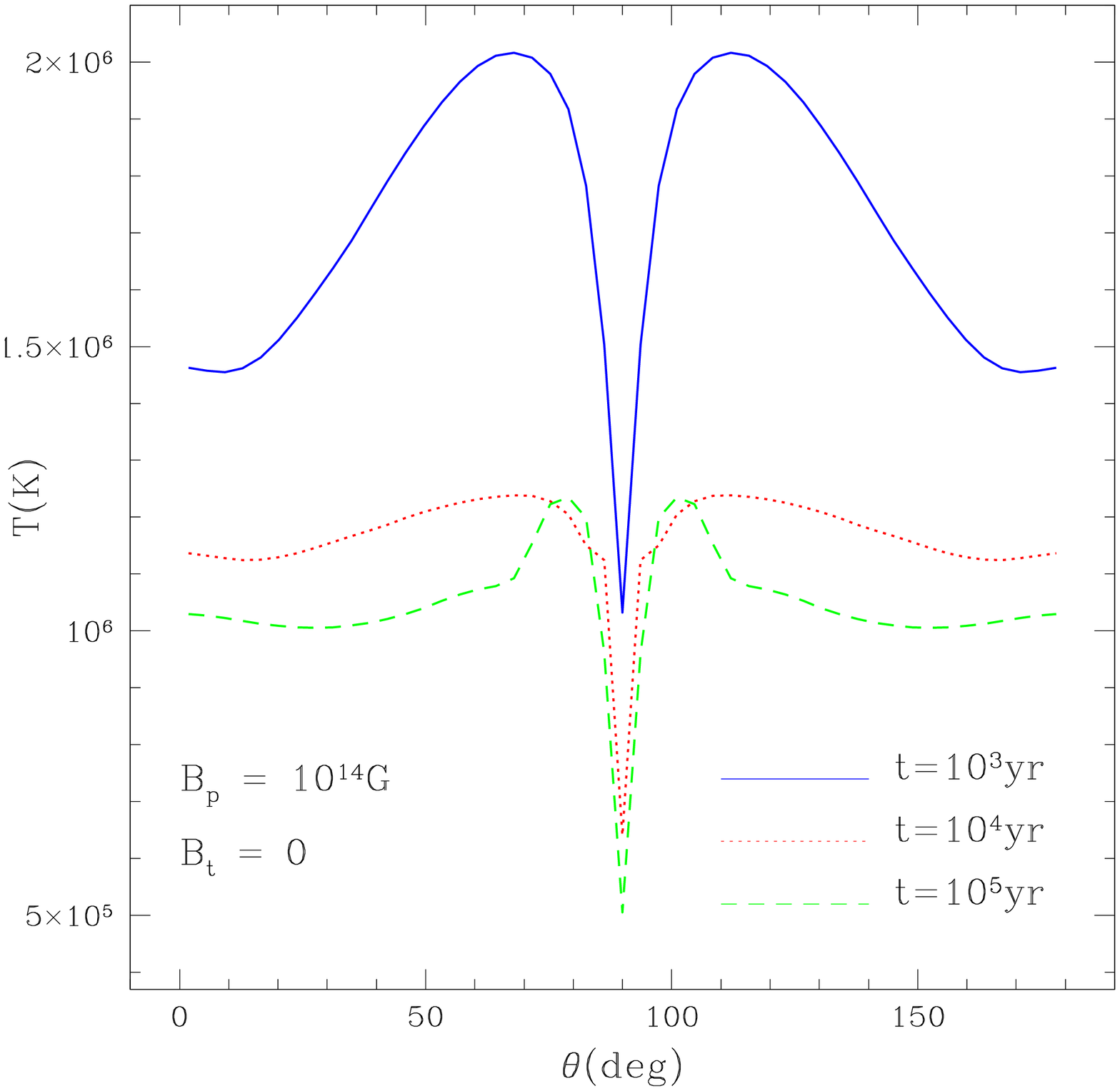}
\caption{Temperature distribution on the surface of the NS, for models
with a purely poloidal field at birth, and at different NS ages.}
 \label{fig:temp1}
\end{figure}

The creation of surface temperature anisotropies in the crust and
envelope of magnetised NSs has been extensively studied by different
groups
\citep{greenstein83,schaaf90,heyl01,potekhin01,geppert04,geppert06,
  perezazorin06a,potekhin07}.  In the outer $\sim$~100 m of a NS, the
temperature gradient in the radial direction is very strong, with a
typical drop of about two orders of magnitude.  In the region where the
magnetic field is radial (e.g., the poles), heat is efficiently
transported along the radial direction, so that the surface is thermally
connected to the inner crust and to the core. On the other hand, the
regions with nearly tangential magnetic field (equatorial area for a
dipolar field) are insulated and thermally disconnected from the hot
core.  Note, however, that 2D models have shown the importance of
tangential heat conduction in limiting the anisotropy, compared with
plane-parallel 1D models.

The degree of anisotropy is controlled by the ratio between thermal
conductivity along and across the field lines, which in a classical
(non-quantizing) approach can be approximated by
\begin{equation}\label{eq:anisotropy}
 \frac{\kappa_\parallel}{\kappa_\perp} \approx 1+(\omega_B\tau_e)^2
\end{equation}
where $\omega_B=eB/m_ec$ and $\tau_e$ are the electron gyrofrequency
and the electron relaxation time (the typical timescale between
scattering processes suffered by electrons), respectively.  { Note
  that the classical approach is a good approximation only for weakly
  quantized matter (see e.g. \citealt{hernquist1984}). For
  magnetar-strength fields, quantum effects become important only at
  low density (the envelope) or low temperature, and the number of
  Landau levels must be consistently calculated. In our simulations,
  we use the full version of quantizing conductivities developed in
  the form of fortran codes by
  A. Potekhin \citep{potekhin99a,potekhin99b,cassisi07,chugunov07},
  which are available online\footnote{
    http://www.ioffe.ru/astro/conduct (we use the 01.02.2013
    version).}.}

At fixed temperature, transport across magnetic field lines is strongly
suppressed for high magnetic field strengths, because $\omega_B
\propto B$, and the anisotropy in the surface temperature is more
pronounced.  However, in some regimes, $\tau_e$ strongly depends on
temperature, so that the so-called magnetization parameter
$\omega_B\tau_e$ varies by several orders of magnitude during the
evolution.  In particular, at low temperatures, $\tau_e$ becomes large
and anisotropy is expected even for weak magnetic fields (see the
discussion in the final part of section 4 of \citealt{pons09}).

The general result for weak dipolar magnetic fields ($B \lesssim 10^{13}$ G) is
that the magnetic poles are systematically hotter than the equatorial region.
This situation can be inverted when internal heating
sources are present. In Fig.~\ref{fig:temp1}, we show examples of the
evolution of the surface temperature for different models parametrized
by the magnetic field strength at the pole ($B_p$).  The top panel
shows results for an initially poloidal, strictly dipolar
configuration, with $B_p=10^{13}$~G. As the star becomes older, the
magnetic field slowly dissipates but the temperature also decreases,
and the surface temperature anisotropy in an old star can sometimes be larger than for a
hot, young star with a higher magnetic field.

The particular temperature profile in the crust
depends on the location of currents and where the dissipated magnetic
energy is deposited.  Eventually, an inversion of the angular
temperature anisotropy in the crust can happen.  However, this effect
is filtered by the envelope: the blanketing effect of the envelope may reestablish the
typical hot poles+cold equator profiles, depending on the strength of
the magnetic field and temperature. In the bottom panel of
Fig.~\ref{fig:temp1}, we show the evolution of the temperature for an
initially poloidal, dipolar configuration, of strength
$B_p=10^{14}$~G. For young NSs, the poles are actually cooler than the
tropical regions, and this counter-intuitive temperature profiles are
kept for longer times for higher fields.

\begin{figure}
\vspace{-0.1in}
\includegraphics[width=.45\textwidth]{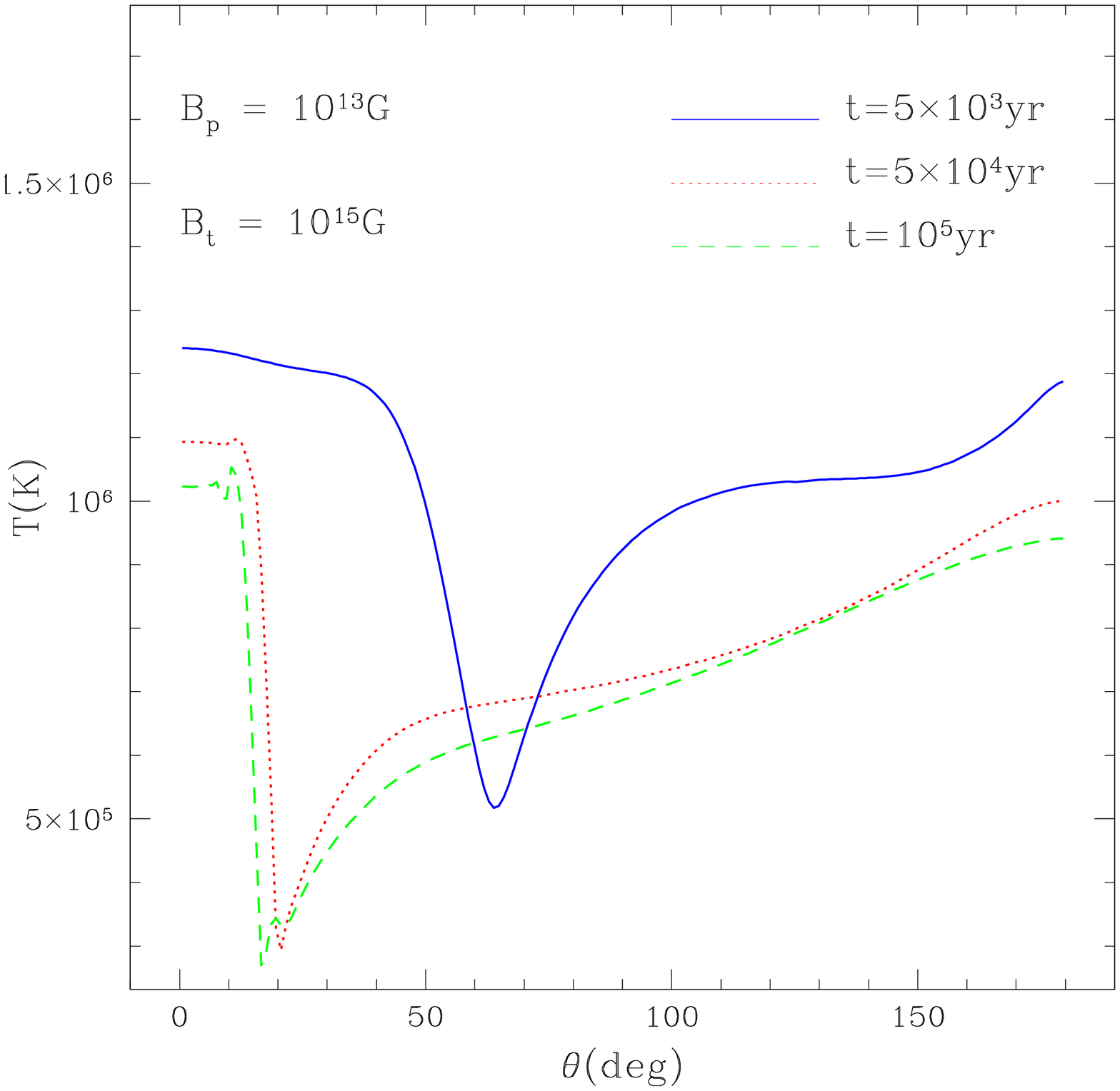}\\
\vspace{-0.1in}
\includegraphics[width=.45\textwidth]{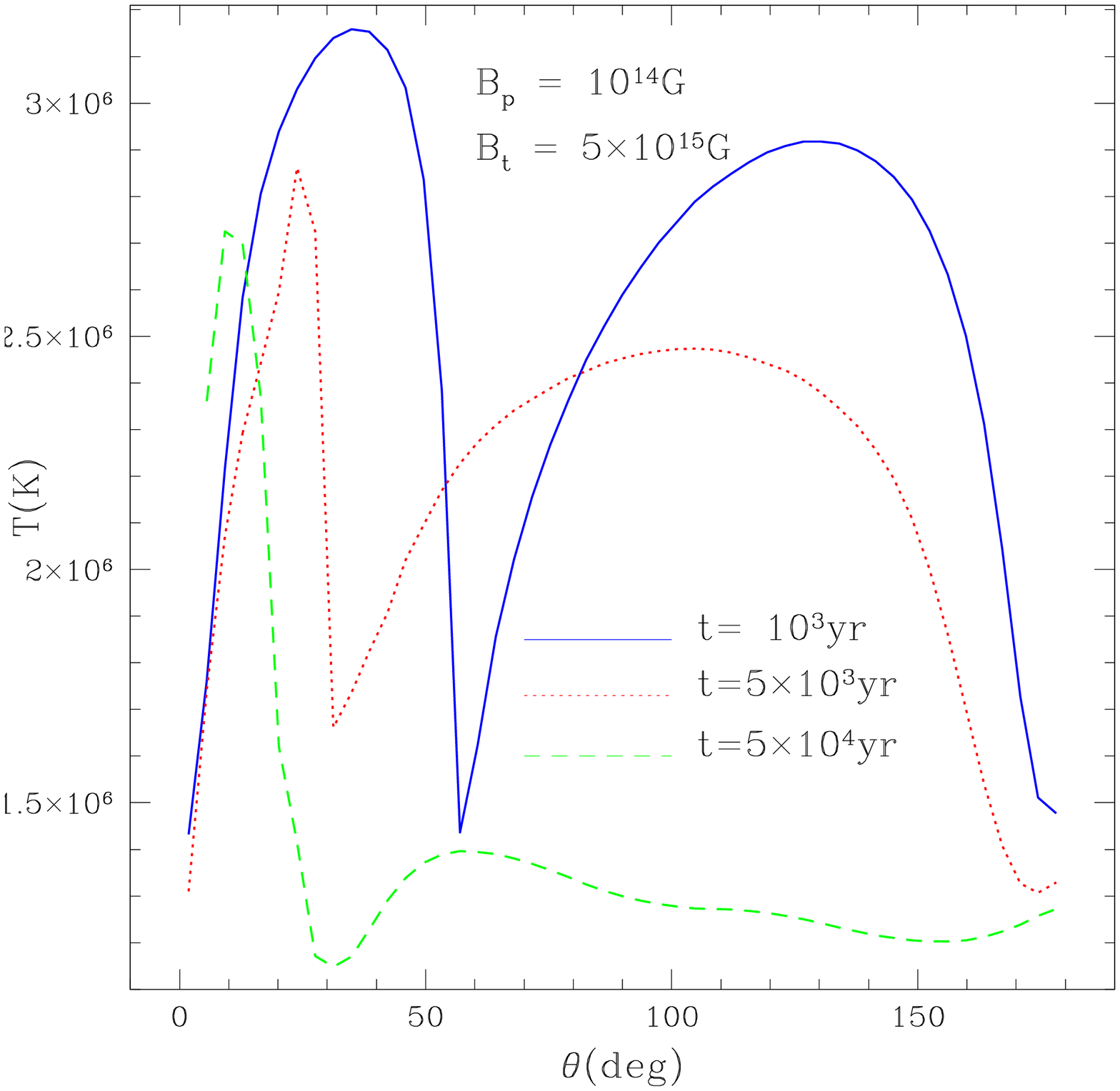}
\caption{Temperature distribution on the surface of the NS, for models
with an initial $B$ field which has both a poloidal and a toroidal component.}
 \label{fig:temp2}
\end{figure}

Note that, in the cases discussed above, the symmetry with respect to
the equator is maintained throughout the evolution, due the initial
choice of a purely poloidal, dipolar magnetic field. However, the
 magnetic field geometry is expected to be much more complicated,
likely with the presence of strong internal components (toroidal field
and poloidal multipoles). In order to explore the importance of 
a strong internal magnetic field, we show in Fig.~\ref{fig:temp2} the temperature profiles
for the same NS models as in Fig.~\ref{fig:temp1}, but with the addition of a strong
toroidal component that contains over $90\%$ of the magnetic energy.

During the evolution, the symmetry is broken due to the Hall term in
the induction equation (see Paper I), which leads to a complex field
geometry with asymmetric north and south hemispheres.  The region with tangential
field lines does not coincide with the equator, and asymmetric
temperature profiles are expected. The degree of anisotropy strongly
depends on the initial toroidal field strength, because of its insulating
effect in the crust. In some cases, a configuration with radial field
lines concentrated at small magnetic latitudes generates a step-like
temperature profile (top panel, $t=10^4,10^5$ yr), with a sharp hot
spot of $\sim 15-20^\circ$. For a given initial magnetic field configuration
at birth, the degree of anisotropy further increases with the age of the 
star. 

 In this paper we assume blackbody emission from the surface, without
 exploring other possible emission models, such as (magnetized)
 atmospheres and a condensed surface. We point out that, under the
 presence of a light-element atmosphere, the emerging spectrum would
 get distorted and, when fitted by a blackbody model, the inferred
 effective temperatures would be a factor $\sim1.5-2$ larger (the
 so-called color-correction factor, e.g. \citealt{suleimanov2011}),
 and, consequently, the inferred radius would be smaller. However, the
 emission model would barely affect the temperature anisotropy.  {
   Furthermore, note that a magnetic atmosphere is expected to result
   in beaming of the outgoing radiation and could affect the observed
   pulse profiles, as widely discussed in the literature
   (e.g. ~\citealt{ozel01}; \citealt{lloyd03}; \citealt{vanadelsberg06};
   \citealt{hoetal07}).  Since here we are specifically interested in
   studying the anisotropy effects (both on spectra and pulse
   profiles) caused by the presence of a crustal field, we work under
   the assumption of local isotropic emission, and briefly discuss
   further possible modifications induced by an atmosphere. } In the
 following, we compute spectra and pulse profiles for the temperature
 distributions of our models.


\section{Spectra and Pulse Profiles.}\label{sec:spectra}

The calculation of the phase-dependent emission from an extended
region on the NS surface follows a well established formalism in the
literature \citep{pechenick83,page95,pavlov00}. In the following, we use the same notation as in \cite{perna12}.
We define the time-dependent rotational phase $\gamma(t)$ as the
azimuthal angle subtended by the magnetic dipole vector
{\boldmath{$\mu$}} around the axis of rotation. It is related to the
modulus of the NS angular velocity, $\Omega(t)$, by
{ $\gamma(t)=\int \Omega(t) dt$}.  The coordinate system is chosen so that the
observer is located along the ${z}$ axis; the inclination angle of the
rotation axis, {\boldmath$\hat{\Omega}$}, with respect to the line of
sight is indicated by $\alpha_R$, while the angle between the magnetic
dipole vector and the rotation angle is denoted by $\alpha_M$. Thus,
the angle between {\boldmath{$\mu$}} and the line of sight is given by
\begin{eqnarray}
\label{eq:alpha}
\cos\alpha(t) = \cos\alpha_R\cos\alpha_M+\sin\alpha_R\sin\alpha_M\cos\gamma(t).
\end{eqnarray}

Due to the strong gravitational field of the NS, the trajectories of photons are
substantially deflected as they travel to the observer. A photon emitted at an
angle $\delta$ with respect to the surface normal will reach the (far away) observer 
if emitted at a colatitude $\theta$ on the star surface, where the relation between
$\theta$ and $\delta$ is given by the ray-tracing function,

\begin{eqnarray}
\theta(\delta) = \int_0^{\frac{R_s}{2R}}du\, x\left[\left(1-{R_s\over R}\right)\left({R_s\over 2R}\right)^2-
               (1-2u)u^2 x^2\right]^{-\frac{1}{2}},
\label{eq:teta}
\end{eqnarray}
\noindent 
where $x\equiv\sin\delta$, and $R_s\equiv 2GM/c^2$ is the Schwarzchild
radius of a star of mass $M$ and radius $R$. Due to the spacetime curvature close to the NS, an observer
at infinity measures a larger NS radius, $R_\infty=R ({1-{\frac{R_s}{R}}})^{-1/2} $.

\begin{table*}
\begin{center}
\begin{tabular}{l c c  c c c c}
\hline
\hline
Model & Age & $T_{eff}$ & $T_{max}$ & $\chi^2$ &  $T_{bb}$ & $R_{bb}$ \\
      & [kyr] & [eV] & [eV] &  &  [eV]   &   [km]   \\
\hline
    B13 &     1 &   113 &   130 &  1.38 &  112$\pm 2$ &     11.9$^{+ 0.6}_{- 0.5}$\\           
    B13 &    10 &    87 &   103 &  0.86 &  93$\pm 1$ &     9.9$\pm 0.4$\\           
    B13 &   100 &    68 &    78 &  1.17 &  72$\pm 1$ &     10.3$\pm 0.5$\\           
    B14 &     1 &   158 &   172 &  1.09 &  162$\pm1$ &   10.8$\pm 0.1$\\ 
    B14 &    10 &   102 &   106 &  1.01 &  100$\pm 1$ &     12.1$\pm 0.5$\\          
    B14 &   100 &    93 &   105 &  0.96 &  95$\pm 1$ &     11.0$\pm 0.4$\\          
   B13t &     5 &    87 &   107 &  0.94  & 92$\pm 1$ &     9.8$^{+ 0.4}_{- 0.3}$\\           
   B13t &    50 &    66 &    94 &  1.19  & 75$\pm 1$ &     8.0$^{+ 0.6}_{- 0.5}$\\           
   B13t &   100 &    63 &    90 &  1.26  &  71$\pm 1$ &     8.2$^{+ 0.6}_{- 0.5}$\\           
   B14t &     1 &   234 &   272 &  1.10  & 241$\pm 3$ &     10.5$\pm 0.3$\\ 
   B14t &     5 &   201 &   255 &  1.11  & 203$\pm 3$ &     11.0$\pm 0.3$\\          
   B14t &    50 &   122 &   246 &  1.56 & 156$\pm 4$ &     6.1$^{+ 0.5}_{- 0.3}$\\          
\hline
\hline
\end{tabular}
\end{center}
\caption{{ Mean and standard deviation of the parameters over 1000
    Monte Carlo realizations} of synthetic absorbed spectra
  ($N_h=10^{21}$ cm$^{-2}$), fitted with the {\tt Xspec} model {\tt
    phabs(zashift*bbodyrad)}.  The theoretical models are the ones
  displayed in Fig.~\ref{fig:temp1} and~\ref{fig:temp2}, labeled by
  the value of $\log(B_p^0)$, where $B_p^0$ is the strength of the
  initial dipolar poloidal field. { The physical radius of the NS
    is 11.6~km in all the models.} The suffix 't' indicates whether
  the initial field is dominated by a strong toroidal component. {
    The reported $\chi^2$ represents the mean value over all the Monte
    Carlo realizations.}}
\label{tab:fits}
\end{table*}

\begin{figure*}
\includegraphics[width=.45\textwidth]{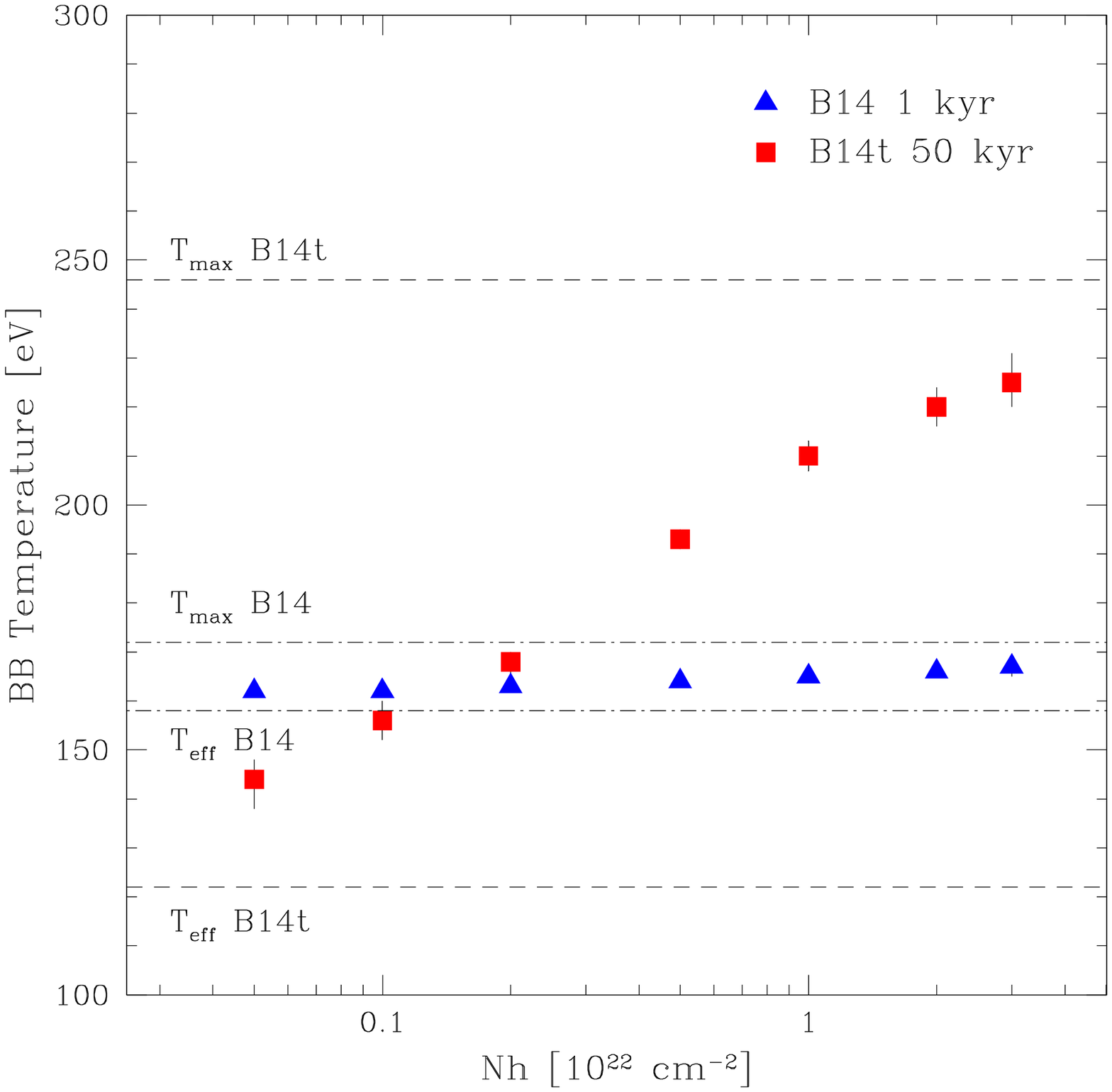}
\includegraphics[width=.45\textwidth]{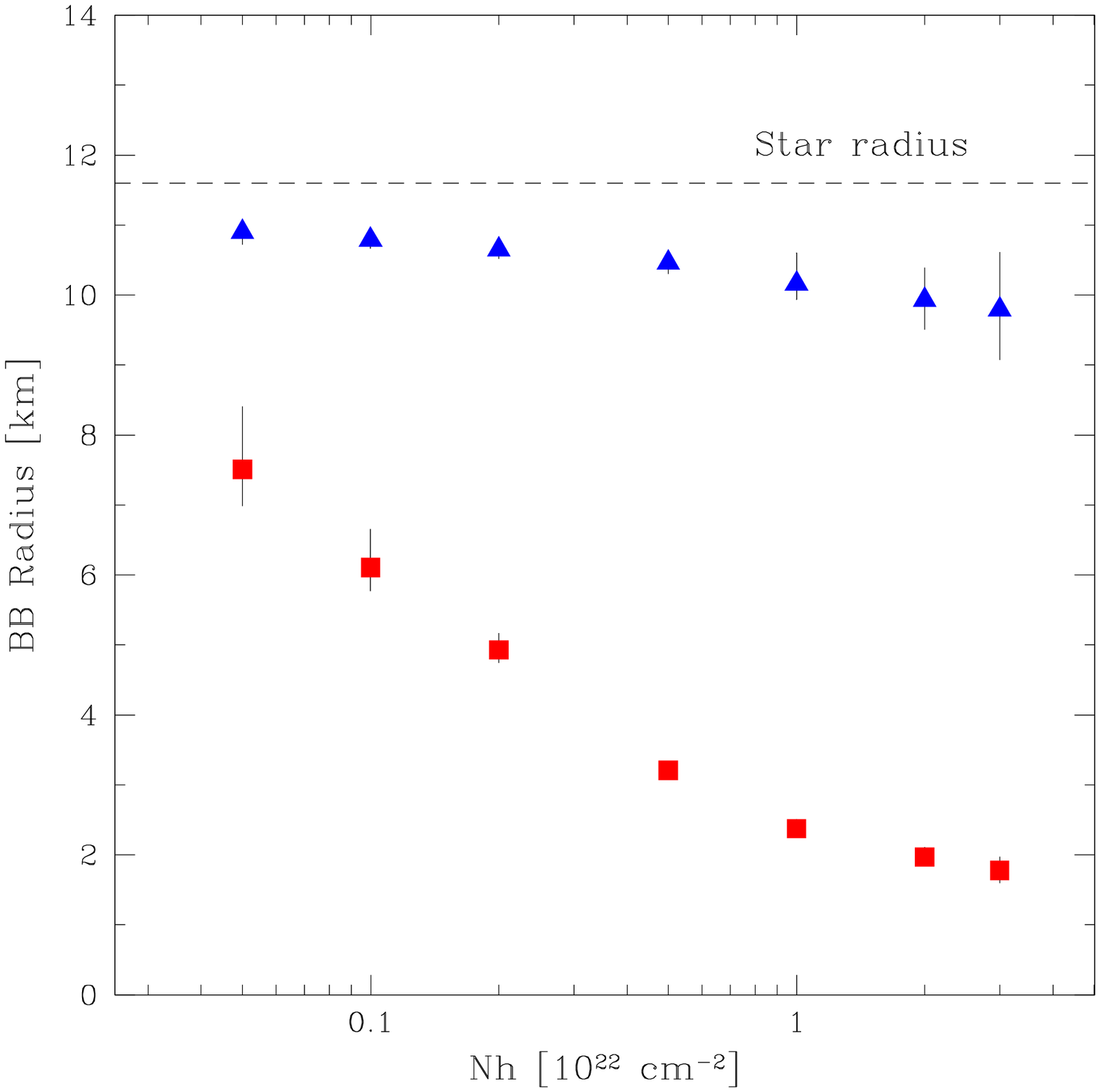}
\caption{Dependence of temperature (left) and radius (right) on $N_h$,
  inferred by the best-fit {\tt phabs(zashift*(bbodyrad))} model for
  two representative models: B14 at 1 kyr (blue triangles) and B14t at
  50 kyr (red squares). Effective and maximum temperatures for both
  models are indicated as references. When a strong toroidal field is
  present (model B14t), the inferred BB temperatures and radii become
  a strong function of the interstellar absorption, unlike the case in which
  the field is purely poloidal (model B14).}
\label{fig:fits_nh}
\end{figure*}

In order to obtain the (phase-dependent) spectrum at the observer one needs to integrate
the local emission over the entire observable surface. Including the effect of gravitational
redshift of the emitted radiation, this integral takes the form
\begin{eqnarray}
F(E_\infty,\alpha) \;=\;\frac{2 \pi}{c^2\,h^3}\;\frac{R_\infty^2}{D^2}\;E_\infty^2
\int_0^1 2xdx\nonumber \\
 \times \int_0^{2\pi}\; \frac{d\phi}{2\pi}\;
I\,\left(\theta,\,\phi,\,E \right)\;,
\label{eq:flux}
\end{eqnarray}
in units of photons~cm$^{-2}$~s$^{-1}$~keV$^{-1}$.
In the above equation, $E_\infty$ is the energy as measured by the
distant observer.  This is related to 
the local photon energy at the stellar surface by $E_\infty = E ({1-{\frac{R_s}{R}}})^{1/2} $.
The function $I(\theta, \phi, E)$ describes the distribution
of the local emission.  For simplicity, here we assume blackbody
radiation at the local $T(\theta, \phi)$. Therefore, the emission is
locally isotropic. This assumption allows us to disentangle the
effects on the light curve modulation due to the temperature
anisotropy (which is what we are interested in),
from those produced by atmospheric effects, which are known to alter,
and generally enhance, the flux modulation, { as mentioned earlier.}

From Eq.~(\ref{eq:flux}), we can easily compute the phase-averaged spectrum as 
\begin{equation}
F_{\rm ave}(E_\infty) = \frac{1}{2\pi}\;\int_0^{2\pi} \;F[E_\infty, \alpha(\gamma)]\; d\gamma\;,
\label{eq:fave}
\end{equation}
as well as the pulse profile in a given (observed) energy band, $\{E_{1,\infty},E_{2,\infty}\}$.
\begin{equation}
F(\gamma) = \int_{E_{1,\infty}}^{E_{2,\infty}}\; F[E_\infty, \alpha(\gamma)] \; dE_{\infty}.
\label{eq:fband}
\end{equation}
The pulsed fraction is defined as 
\begin{equation}
{\rm PF} = \frac{F_{\rm max}(\gamma)- F_{\rm min}(\gamma)}{F_{\rm max}(\gamma)+ F_{\rm min}(\gamma)}\;,
\label{eq:pf}
\end{equation}
where the phases corresponding to the maximum and minimum of the flux
may vary depending on the temperature distribution on the NS surface,
{ as well as on the chosen energy band}. With the local thermal emission
assumed to be blackbody, phase-averaged spectra are { multitemperature}
blackbodies.

\subsection{Synthetic spectra and blackbody fits}

Before studying the effect of the temperature anisotropy on the
observable pulse profiles, we wish to address another related question of
great importance for the interpretation of observations:
given a spectrum from a realistic temperature distribution, such
as those displayed in Figs.~\ref{fig:temp1} and \ref{fig:temp2}, what are the
NS radius and temperature that would be obtained with a blackbody fit
to the thermal component, as routinely done? 

To answer this question, we imported the phase-averaged spectra from
the temperature distributions in Figs.~\ref{fig:temp1}, \ref{fig:temp2}
into the fitting software {\tt Xspec} \citep{arnaud96}. Using the
recently developed {\tt flx2tab} function in the {\tt \small HEASARC}
package, v6.13, we created the {\tt atable} models for each one of the
spectra. With {\tt Xspec}, we then simulated synthetic observed
spectra, using {\em XMM-Newton}/EPIC-pn response matrices in the {\tt
  fakeit} procedure. We simulated also the interstellar absorption, by
means of the {\tt phabs} photoelectric absorption
model with the \cite{anders89} abundances, and the \cite{balucinska92}
photoelectric cross-sections. 
 For all the models we discuss in
this paper, the normalization is fixed by assuming a
distance of 1 kpc, and the exposure time of the simulated observation
is chosen so that the total number of photon counts would be of
$\approx 5000$ counts, as typical of realistic observations. 
{ For each model, we simulated 1000 random realizations of the 
spectrum.}
Then, { for each spectrum}, we proceeded to the spectral analysis
in the standard way. In real observations, other contributions
(e.g. hard tails, non-thermal components from rotation) extend the
energy range to higher energy, but here we focus on the thermal
contribution alone. We fitted the simulated spectra with the model
{\tt phabs(zashift*blackbody)}, where {\tt zashift} applies the
redshift correction, and {\tt phabs} models the interstellar
absorption.  Our theoretical models for the temperature profiles and
spectra were computed with $M=1.4M_\odot$ and $R=11.6$~km (same NS
model used in the simulations), yielding a redshift of
$z=0.25$.\footnote{Note that, without redshift corrections, we would
  obtain exactly the same fit, but with temperature and radius at
  infinity, i.e. $T^\infty=T/(1+z)$ and $R^\infty=(1+z)R$.}  We
included an interstellar absorption of magnitude $N_h=10^{21}$
cm$^{-2}$, which is typical of nearby sources.  The
results of the fits for all the models considered here are reported in
Table~\ref{tab:fits}. { For the fitted parameters ($T_{bb},
  R_{bb}$), we report the mean over the 1000 realizations, as as well
  the standard deviation.}

In general, the temperature inferred by a single blackbody fit {
  lies} between the effective temperature $T_{\rm eff}\equiv (L_{\rm
  bol}/4\pi\sigma R^2)^{1/4}$ and the maximum temperature of the
model, both shown in the table as well.  By looking at the table and
at the temperature profiles, we can roughly distinguish two kinds of
models.  In most models with a dominating poloidal magnetic field
(B13, B14), the cold regions located near the equator have a small
angular size and barely contribute to the total flux. Hence, the value
of the inferred radius is only slightly smaller than the star radius,
and the temperature is only slightly higher than the effective
temperature, which is not very different than the maximum
temperature. On the other hand, when a strong toroidal component is
present, the temperature profile, especially at old ages, can be
roughly described by a cold component from a large part of the surface
and a smaller, hot region. In these cases, the inferred BB radius
(whereas a single BB provides a good fit to the thermal component) can
be much smaller than the actual NS radius. The amount of reduction is,
as it may be expected, a strong function of the energy-dependent
absorption by the interstellar medium.

To study this effect, we analyzed in detail two representative models:
B14 at 1 kyr and B14t at 50 kyr. For each of them, we generated
synthetic spectra for values of absorption in the range
$N_h\in[5\times 10^{20} , 2\times10^{22}]$ cm$^{-2}$. Given the
typical temperatures of our models,
larger values of $N_h$ would make the thermal emission from the source
undetectable. In Fig.~\ref{fig:fits_nh}, we show the dependence of the
best-fit blackbody parameters (radius and temperature) on $N_h$ for
both models.  For the B14 model at 1 kyr (blue triangles), the best-fit
temperatures and radii are almost independent of $N_h$, with only a
slight overestimation of temperature (and underestimation of radius)
for the strongest absorption. For $N_h \lesssim 10^{22}$ cm$^{-2}$,
the inferred temperatures are within the 10\% variability between the
effective and maximum temperature, and the radius is compatible with,
or a little smaller than the real radius. On the other hand, model
B14t at 50 kyr (red squares) shows a large variability in the inferred temperature,
caused by the large anisotropy and larger range of temperatures on the
star surface. The inferred temperature tends to approach the
highest temperature only for large absorption values, while it becomes
lower as $N_h$ decreases. The inferred radius of the emitting region
is always significantly smaller than the real star radius, and can
vary by up to a factor of four depending on the value of $N_h$, for
the same model.

\begin{figure*}[ht]
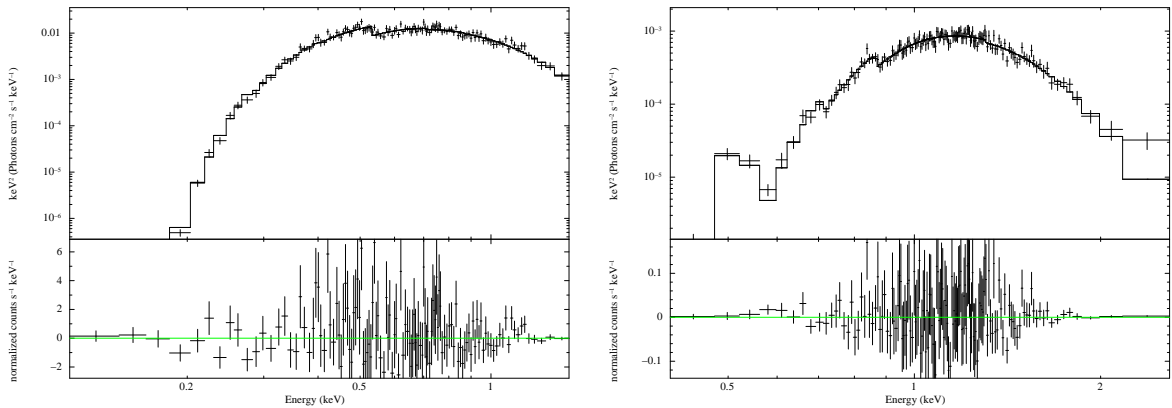

\includegraphics[width=.3\textwidth,angle=270]{b14y21.ps}
\includegraphics[width=.3\textwidth,angle=270]{b14y22.ps}
\caption{Synthetic spectra and best-fit blackbody model and residuals
  for the B14 model at 1 kyr, with $N_h=10^{21}$ cm$^{-2}$ (left panel)
  and $N_h=10^{22}$ cm$^{-2}$ (right).}
\label{fig:fits_spectra_b14y}
\end{figure*} 

\begin{figure*}[ht]
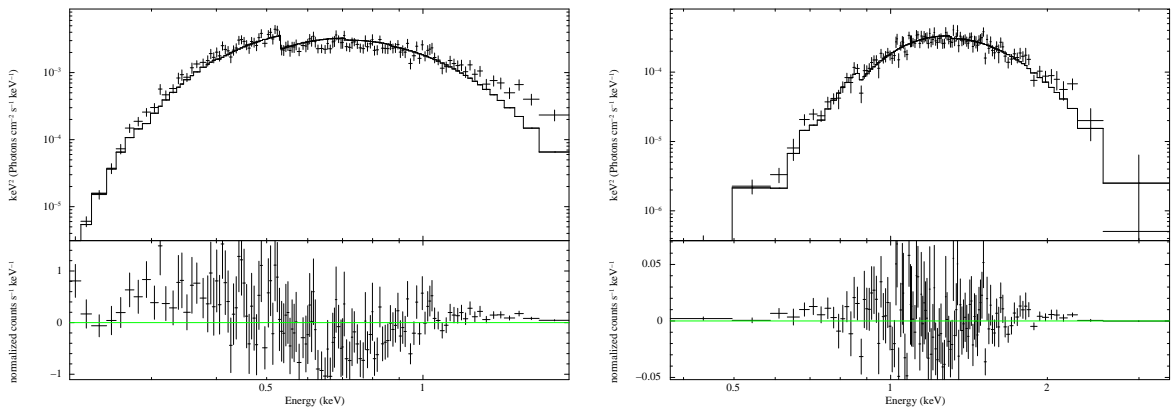

\includegraphics[width=.3\textwidth,angle=270]{b14to21.ps}
\includegraphics[width=.3\textwidth,angle=270]{b14to22.ps}
\caption{Synthetic spectra and best-fit blackbody model and residuals
  for the B14t model at 50 kyr, with $N_h=10^{21}$ cm$^{-2}$ (left panel)
  and $N_h=10^{22}$ cm$^{-2}$ (right).}
\label{fig:fits_spectra_b14to}
\end{figure*} 

In Fig.~\ref{fig:fits_spectra_b14y} we show the simulated spectrum and the
best-fit blackbody { in one realization of the B14} model at 1~kyr,
for two different values of the absorbing column density, $N_h
=10^{21}$~cm$^{-2}$ and $N_h =10^{22}$~cm$^{-2}$. In both cases, the
single blackbody fit is acceptable and there is no evidence of
anisotropy (i.e. need for additional components) in the spectrum.

For low values of absorption, the fake spectrum of model B14t at 50
kyr cannot be fitted accurately by a single blackbody. In the left
panel of Fig.~\ref{fig:fits_spectra_b14to}, we illustrate { one
  realization} for $N_h=10^{21}$ cm$^{-2}$, which is not acceptable
($\chi^2 \sim 1.6$).  In this case, a much
better fit ($\chi^2=1.2$) is provided by two blackbodies with 
$kT_1\simeq 90$~eV, $kT_2\simeq 165$~eV, $R_2\sim 5$~km,  and $R_1$
compatible with the star radius, reflecting the real temperature
profile. Larger values of
absorption ($N_h=10^{22}$ cm$^{-2}$ in the right panel of
Fig.~\ref{fig:fits_spectra_b14to}) ``hide'' the cold component, so
that the fit provides an estimate of the temperature and size of the
hotter component, much smaller than the real radius. This explains the
trend of Fig.~\ref{fig:fits_nh} (right panel) with absorption:
  the larger the absorption, the hotter is the effective BB
  temperature measured from the fit, and hence the smaller is the
  inferred radius of the emitting region.

\subsection{Pulse profiles}\label{sec:pulse}

\begin{figure}
\vspace{-0.1in}
\includegraphics[width=.45\textwidth]{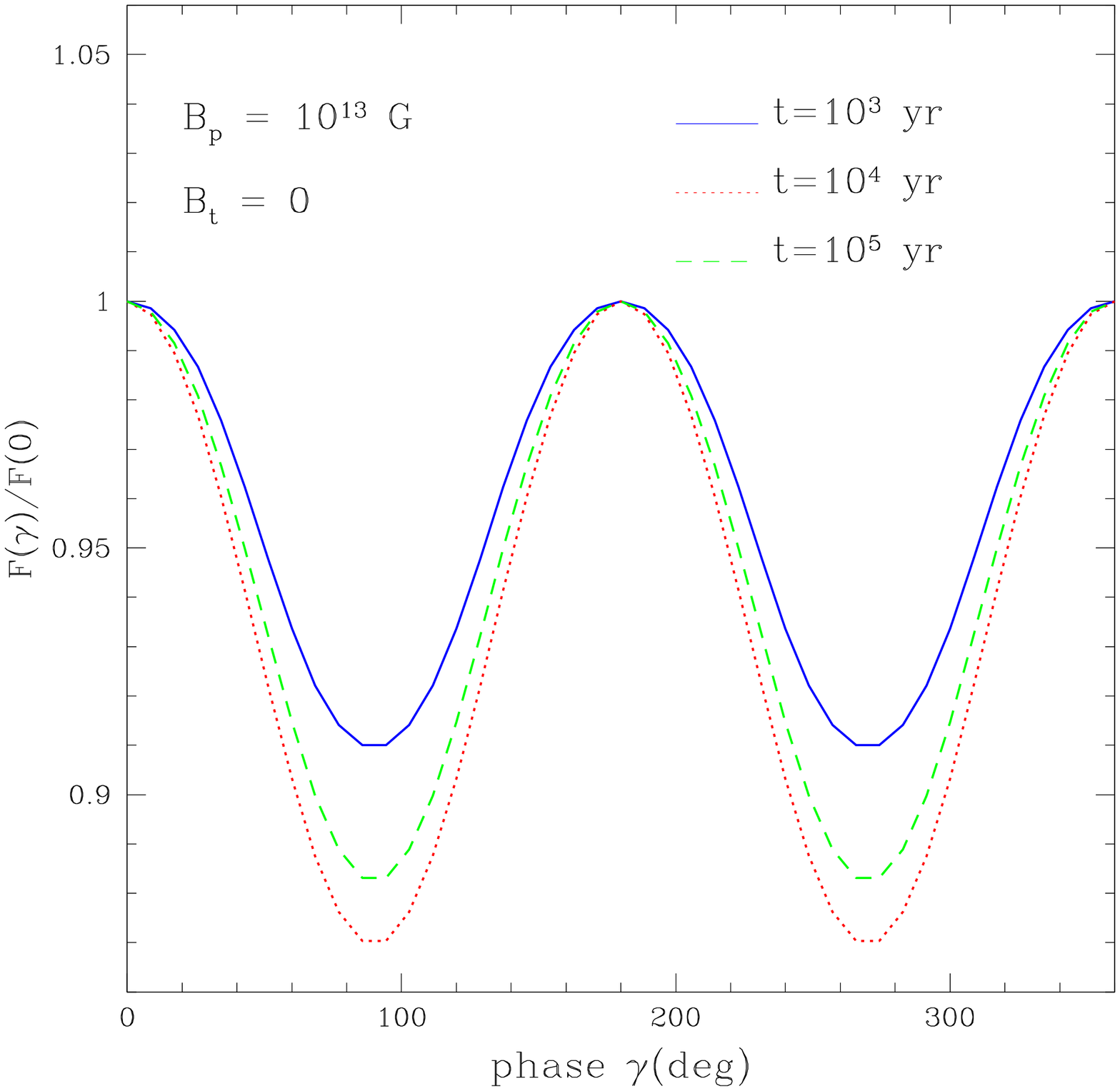}\\
\vspace{-0.1in}
\includegraphics[width=.45\textwidth]{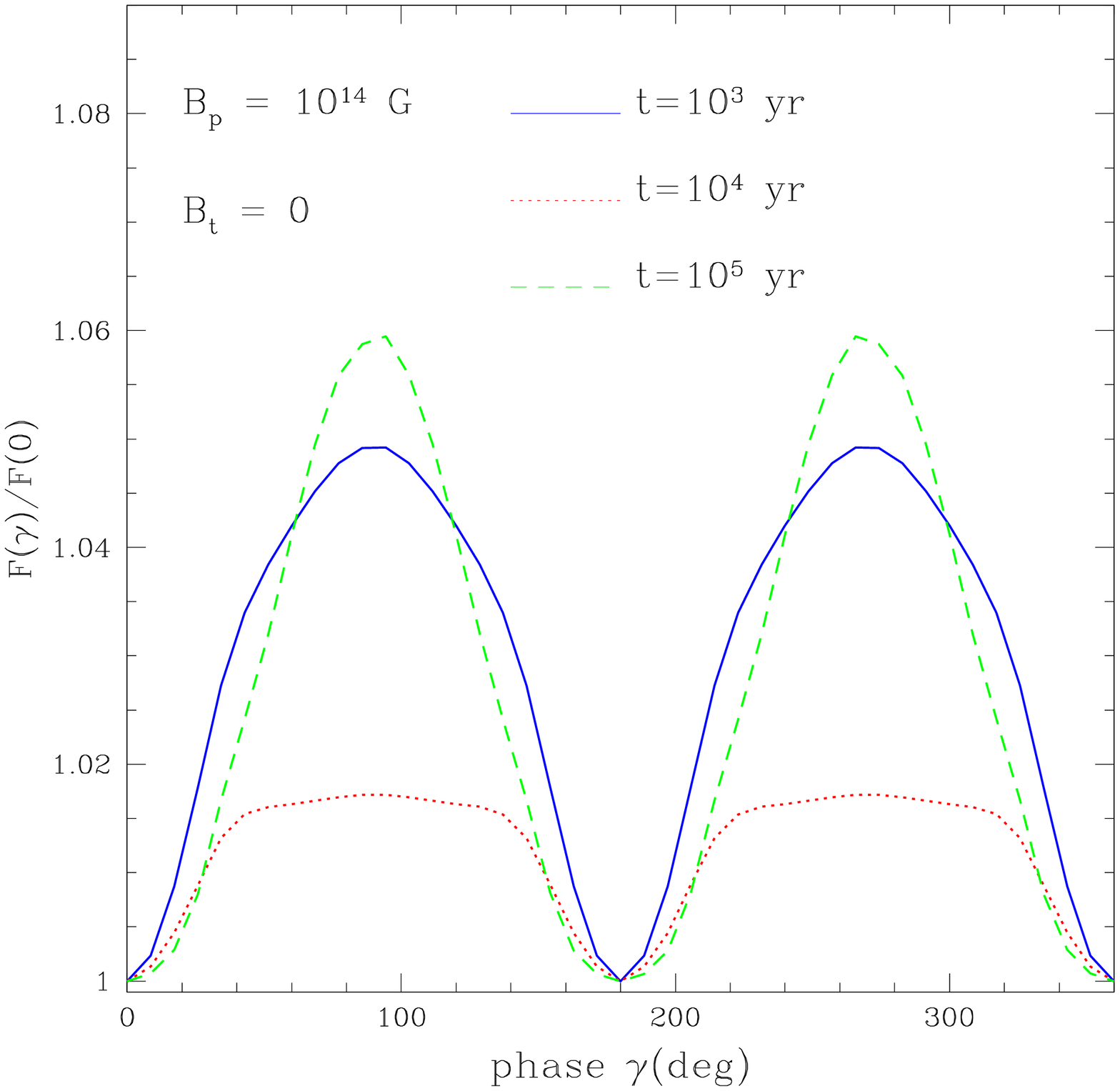}
\caption{Pulse profiles in the 0.5-2~keV band for the temperature
  profiles of Fig.~\ref{fig:temp1}.} 
 \label{fig:lc1}
\end{figure}

\begin{figure}
\vspace{-0.1in}
\includegraphics[width=.45\textwidth]{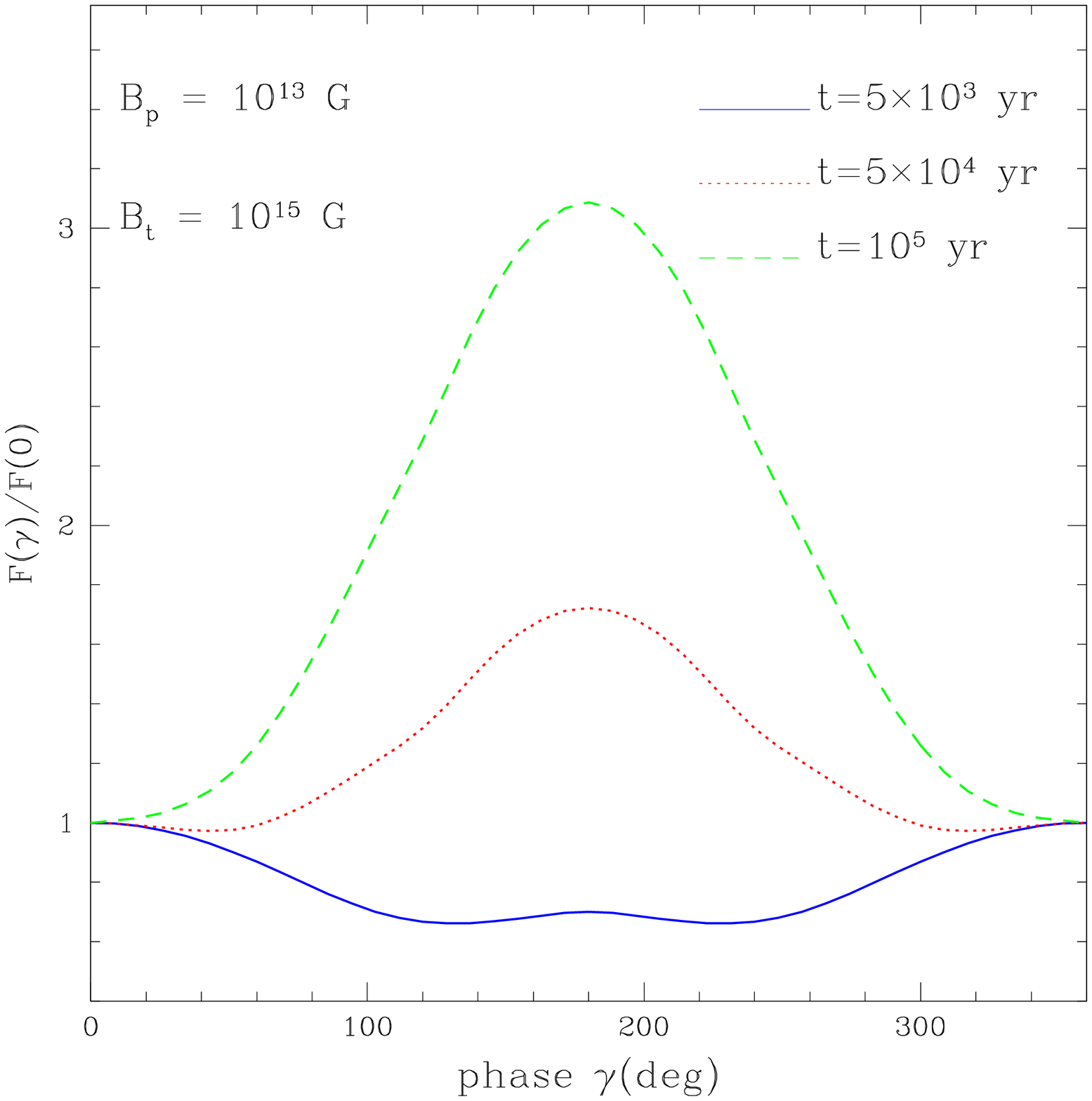}\\
\vspace{-0.1in}
\includegraphics[width=.45\textwidth]{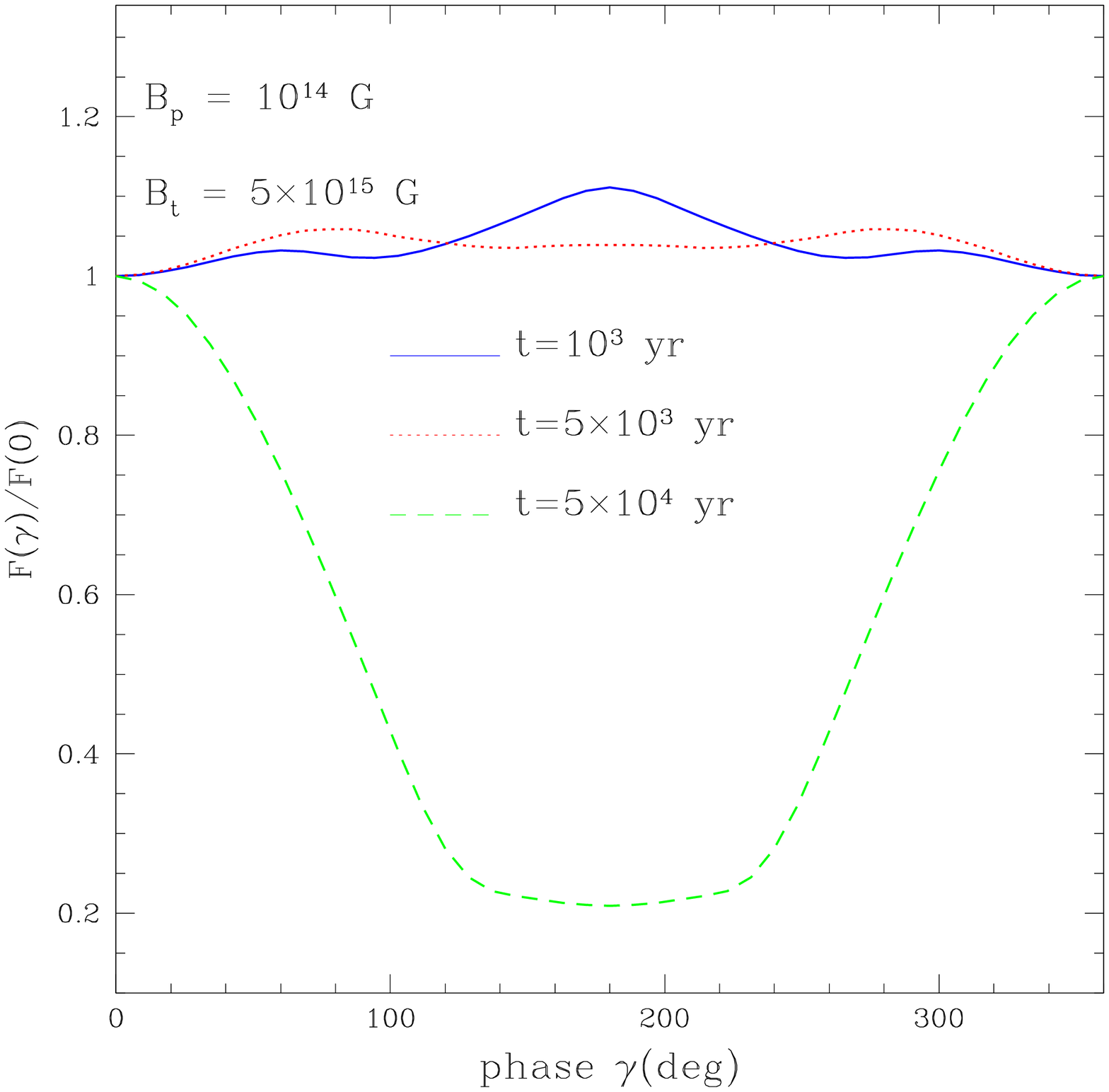}
\caption{Pulse profiles in the 0.5-2~keV band for the temperature
profiles of Fig.~\ref{fig:temp2}.}
 \label{fig:lc2}
\end{figure}

As evident from the discussion of the previous section, spectral
modeling is generally degenerate: when the resolution is not extremely
high, several models often fit equally well the same spectrum. This
has been especially demonstrated by the results in
Table~\ref{tab:fits}: while the intrinsic spectra are composite
blackbodies with a variety of temperature profiles, the resulting
spectra can generally be fit by a single blackbody.  A much deeper
probe of the intrinsic temperature profile of the NS surface is
provided by the pulse profile $F(\gamma)$ (cfr Eq.\ref{eq:fband}) of
the star in a thermal energy band. This is displayed in
Figs.~\ref{fig:lc1} and \ref{fig:lc2} for the set of models presented
in Figs.~\ref{fig:temp1} and \ref{fig:temp2}.  Fluxes are integrated
over the typical soft X-ray [0.5-2]~keV band.  While the thermal flux
generally dominates at lower energies, in practice, a combination of
typical instrumental sensitivity and high interstellar absorption
makes this band more relevant for comparison with observations.  Note that
the observed pulse profiles depend on the viewing angle (parametrized
by $\alpha_M$ and $\alpha_R$);  here we consider, as our base model, the
simplest geometry with $\alpha_R=\alpha_M=90^\circ$ (orthogonal
rotator).  For this configuration, the viewing angle $\alpha$ becomes
equal to the phase angle $\gamma$.

Some general features can be immediately inferred from inspection of
Figs.~\ref{fig:lc1} and \ref{fig:lc2}. When the field is purely
poloidal and hence the temperature profile is symmetric with respect
to the equator (cfr. Fig.~\ref{fig:temp1}), the pulse profile is
double peaked (Fig.~\ref{fig:lc1}).  However, the maximum of the
emission may or may not coincide with the sweeping of the axis of the
magnetic dipole across the line of sight. This is a consequence of the
fact that, for moderate strengths of the poloidal field, the
temperature is higher at the poles (top panel of Fig.~\ref{fig:temp1},
$B_p=10^{13}$~G), hence resulting in a flux maximum corresponding at
$\gamma=0$, while for stronger fields (bottom panel of
Fig.~\ref{fig:temp1}), the hotter regions shift at intermediate
latitudes on the NS surface, hence producing a maximum of the
pulsation at large viewing angles.

A common feature of all the models with poloidal field alone (assumed
purely dipolar) is that the pulsed fraction is very low; for a local
isotropic photon distribution as assumed in Fig.~\ref{fig:lc1}, it
remains constrained to within a few percent.  This is the maximum that
can be obtained without other effects, since we have considered an
orthogonal rotator, which maximizes the flux variations. For
geometrical configurations with $\alpha_R$ and/or $\alpha_M$ smaller
than $90^\circ$, pulsed profiles become asymmetrical, and the pulsed
fraction becomes even smaller.  However, there are a number of effects
which can increase the magnitude of the pulsed flux.

Firstly, high interstellar absorption, by preferentially absorbing low
energy photons with respect to the high energy ones (more pulsed for
BB emission), tends to increase the overall pulsed fraction of the
thermal component in a wide energy band. To estimate the magnitude of
the effect, we ran a few simulations with a column density
$N_H=10^{22}$~cm$^{-2}$, and using the absorption cross sections by
\cite{morrison83}.  For a typical case with $B_p=10^{13}$~G and
$t=10^4$~yr, we found that the amplitude of the pulsed fraction
increased from 3.5\% to 6.1\% in the 0.5-2~keV band.  Other (purely
poloidal) cases are similar. A larger increase in the modulation can
be typically obtained when considering wider energy bands
(e.g. 0.05-10~keV), or larger absorbing column densities, as
extensively discussed in \cite{perna00}.  However, for typical
observational energy bands and absorbing columns, the PF of models
with a mainly dipolar poloidal field remains small.

A more significant influence on the flux modulation derives from
anisotropy in the local emission (see e.g. the parametric study by
\citealt{dedeo01}). Magnetized atmospheres are known to create
anisotropic radiation patterns (e.g. \citealt{vanadelsberg06};
\citealt{ho08}). A detailed computation of lightcurves with realistic,
magnetized atmosphere profiles is beyond the scope of this work, as it
would require a large number of atmosphere models, each for a certain
temperature and varying magnetic field inclination over the surface of
the star. Atmosphere models with inclined fields
\citep{pavlov94,zavlin95,lloyd03,ho07} are computationally very
expensive\footnote{An approximate method for computing spectra and
  light curves with arbitrary temperatures and magnetic fields has
  been developed by \citet{shabaltas12}.}. Hence, following a number
of previous investigations { \citep{perna00,dedeo01,perna08}}, we
consider a parametrized form for the local radiation beaming, which we
implement numerically by weighing the intensity $I$ in
Eq.~(\ref{eq:flux}) by a function $f(\delta)$, where $\delta$ is the
angle between the normal to the surface and the direction of photon
propagation.  { Note that, in realistic models of magnetized
  atmospheres, the shape of the radiation beam is a function of $B, E,
  T_{\rm eff}$.  For magnetic fields normal to the NS surface,
  \citet{vanadelsberg06} find that there is a narrow pencil beam
  centered around the B field direction, accompanied by a broader fan
  beam at larger angles. However, for magnetar-level fields, and with
  the inclusion of vacuum polarization effects, the gap between the
  two beams is reduced, and the radiation emerges as a broad,
  featureless pencil beam centered along the B field direction. For
  inclined B fields, the pencil component of the beam still follows
  the direction of the field (e.g. \citealt{pavlov94,zavlin95,ho07}).}

In the following, with the purpose of simply providing an indication
of the effect of a local anisotropic emission on the observed PFs of a
purely dipolar, poloidal magnetic field configuration, { we assume
  pencil beaming from a magnetic field normal to the surface, which we
  model as $f(\delta)\propto\cos^n\delta$, and} we report the value of
the PFs from pulse profiles computed with a beaming intensity $n=1$.
We again focus on the 0.5-2~keV band and, for each temperature
distribution of Fig.~\ref{fig:temp1}, we report below the results for
the model at the time $t$ at which the PF is the highest (best
scenario for obtaining a highly modulated flux).  We found that the PF
of the $B_p=10^{13}$~G, $t=10^4$~yr model, increases from the 6.9\%
value of the isotropic case to 25\% of the $n=1$ case.  In the case
with $B_p=10^{14}$~G, $t=10^5$~yr, the PF increases from 2.8\% to
12.2\%.  Clearly, a more extreme beaming of the local radiation would
further increase the PF \citep{dedeo01,perna00}. Overall, our results
show that a predominantly poloidal (dipolar) field results in double
peaked profiles with a low level of flux modulation, unless the local
emission is highly anisotropic.

These considerations change dramatically when the field has a strong
large-scale toroidal component. The evolution of an initially strong
toroidal, dipolar field breaks the symmetry of the temperature profile
with respect to the equator (cfr. Fig.~\ref{fig:temp2}); an
interesting consequence is that the resulting light curve becomes {\em
  single peaked}, and the pulsed fraction is much higher, reaching
over 50\% in the { 0.5-2~keV} band for some of the models (see
Fig.~\ref{fig:lc2}). This is a particularly important result: without
appealing to magnetospheric or beaming effects, purely thermal models
{\em can} explain large pulsed fractions and single pulse profiles
observed in a number of isolated NSs.   We stress that this effect
  is achieved if the magnetic field has a strong large-scale component
  tangential to the surface (e.g., in the polar or azimuthal
  direction), which acts as an insulator for a large fraction of the
  surface. Other models, with high multipolar components, yield more
  complicated temperature profiles, with multiple small cold and hot
  regions. The resulting pulsed fraction, which averages out the
  small-scale differences, is small.

\section{Discussion}\label{sec:obs}

 As our simulations show, the surface temperature (and hence the X-ray
 pulsed profile) of a NS changes dramatically with its initial field
 strength and topology, as well as its age.  The theoretical spectra
 and pulse profiles presented in the previous section bear direct
 implications for the observations of X-ray emission from isolated
 NSs. A direct comparison between our numerical results and X-ray
 spectra and pulse profiles of specific objects cannot be made unless
 the simulations are specifically tailored to particular objects,
 which is beyond the scope of this work. Hence, in the following, we will
 discuss some general features within the context of our findings.

\subsection{Magnetars.}

Magnetar spectra in the 0.3-10~keV energy range are typically best fit
by the combination of a blackbody and a powerlaw. The thermal
component, which often dominates in the lowest energy band, can be
generally accounted for by a single BB; the inferred emission regions
(for the best estimated distances) are most often found to be smaller
than the whole surface of the NS. In fact, the small emitting radii
appeared to argue in favor of alternative models, such as accretion
from a fallback disk (e.g. \citealt{chatterjee00}), despite being
ameliorated by fits with atmosphere models rather than pure BB ones
\citep{perna01}.

Our magnetothermal models have shown that certain magnetic
configurations (and especially some of those with a strong large-scale
toroidal component) can result in inferred radii apparently
incompatible with emission from the entire surface of the star when
fitted with a pure blackbody, without atmospheric corrections.  In
particular, among the examples of theoretical models presented here,
the one indicated with B14t at 50~kyr is especially illustrative.  At
small or moderate levels of interstellar absorption ($N_h\lesssim
5\times 10^{21}$~cm$^{-2}$), this model is best fit by a
two-blackbody: one cooler component consistent with emission from the
entire surface of the star, and a hotter component with a smaller radius
of the emitting region ($\sim$ 5~km).  However, for more significant
values of interstellar absorption ($N_h\gtrsim 
10^{22}$~cm$^{-1}$), the cooler component is hidden, and then the NS
appears as emitting only from a single, small, 'hot spot' of about  1-2~km
in size { (see also results by \citet{shabaltas12} with
  analytical profiles for the magnetic field)}.  This is indeed often the case
for magnetars which, interestingly, have absorption column densities
that typically exceed $\sim 10^{22}$~cm$^{-2}$
\citep{esposito08,rea05,gelfand07}.

Thermal pulsed profiles of magnetars display a mixed morphology. A
notable example of a (almost symmetric) double peaked light curve is
the magnetar 1E~2259+586 \citep{patel01,woods04}.  A double peaked
profile could be obtained with an initial poloidal, dipolar field, and
a toroidal, quadrupolar configuration.  However, the high PF at about
the 20\% level of the thermal flux could only be accounted for by a
significant anisotropy in the local emission ($f(\delta)\propto
\cos^{1.5}\delta$ in the notation of \S~\ref{sec:spectra}).  Another
double peaked magnetar is 4U~0142+0162; the low level of modulation of
this object (around the 10\% level) does not constitute a problem for
symmetric temperature profiles \citep{rea07}. Also for this object,
the dominant toroidal component would have to be of even parity in
order not to break the symmetry with respect to the equator.

For most other magnetars, on the other hand, the pulsed thermal
(quiescent) component is single peaked and often highly
pulsed. Examples are\footnote{ In the quoted references, when the
  thermal band is not explicitly separated, we simply refer at the
  pulsed profile in the lowest energy band.} (among others)
1E~1048.1-5937 \citep{tam08}, XTE~J1810-197
\citep{bernardini09,bernardini11}, 1E~1207.4-5209 \citep{halpern11},
1E~1547.0-5408 \citep{dib12}, SGR~0418+5729 \citep{rea13},
SGR~J1822.3-1606 \citep{rea12}.  The single peaked, highly modulated
flux is generally associated to emitting areas that are
much smaller than the entire surface of the NS. Single peaks coupled with a
large PF rule out symmetrical emission geometries with small angles
$\alpha_R$ and/or $\alpha_M$. They rather imply a temperature
distribution on the surface of the star which is asymmetric with
respect to the equator.

Our results hence suggest that, for these objects, a strong toroidal
component must be present in the interior of the star. We note that,
in particular for the case of the low-$B$ magnetars SGR~0418+5729 and
SGR~J1822.3-1606, a strong internal field has been invoked for the
objects to have a non-negligible probability of an outburst
\citep{rea12,rea13}.  More generally, it has been discussed
(\citealt{perna11}; \citealt{pons11}; Paper I) how a strong
internal toroidal field is more likely to make a NS to appear as a
magnetar (i.e. higher thermal luminosity, higher outburst frequency)
as compared to a NS with a similar dipolar field but insignificant
toroidal composnent.  Our findings about the shape of the pulse
profiles, as well as on the sizes of the inferred blackbody radii,
provide an additional support to the earlier suggestions.

We note, however, that tiny spots (a
fraction of km in size), such as, for example, the one measured for
the thermal emission of CXO~J164710.2-455216
\citep{muno06,skinner06,israel07} would be hard to explain with
anisotropic, internal heat alone. Such intensely heated, very small
regions are more likely to result from currents in a twisted magnetic
bundle \citep{beloborodov09,turolla11}.

\subsection{The magnificent seven.}

Another class of objects for which our results are of direct relevance
is that of the X-ray Isolated Neutron Stars (XINSs, also known as the
``Magnificent Seven'', { and first discovered by
  \citealt{walter96}}). These are characterized by purely thermal
spectra, blackbody radii of a few km in size, and generally modest
pulsation levels, ranging from a few percent to 18\%. These objects,
with inferred dipolar fields in the few $\times 10^{13}$~G range, have
timing properties and X-ray luminosities consistent with those of
evolved magnetars\footnote{The birthrates of magnetars and of XINSs
  have been computed by a number of
  authors~\citep{gill07,keane08,ferrario08,popov10}; however, the
  inferred numbers are still rather uncertain due to small number
  statistics and the recent discovery of low field
  magnetars \citep{rea10}, as well as of transient magnetars
  \citep{ibrahim04}.} {(for a review of their properties see e.g.
  \citealt{trumper05}; \citealt{turolla09})}, i.e. stars born with
dipolar fields of a few $10^{14}$~G, which decayed to the current
value within a few $\times 10^5$~years { (see Paper I for a thorough
  description of the coupled evolution of magnetic field and
  temperature).}  The observed pulse profiles are close to sinusoidal,
and single-peaked except for RX~J1308.6+2127, which is double-peaked.

Consistently with previous work
\citep{page95}, which used simplified, analytical models for the
dipolar temperature distribution, we find that single peaked profiles
from purely dipolar, poloidal fields can only be produced for a small
range of oblique viewing geometries and, when so, they would yield
very low PFs, typically not exceeding a few percent if the local
emission is purely isotropic. Anisotropic emission can largely
increase the PF as discussed in Sec~3.2, and previous work
\citep{zane06} has shown that it is possible to fit reasonably well
the pulsed profiles of these objects with a combination of Hydrogen
atmospheric models, and a star-centred dipole plus a quadrupole
topology.

\subsection{Central Compact Objects.}

Last, we conclude with the discussion of another class of isolated NSs
whose observational properties can be interpreted within the context
of our findings -- that is the sample of objects known as central
compact objects or CCOs. These relatively young objects (often still
found at the center of their supernova remnant) are characterized by a
low inferred external (dipolar) field, generally $B_p\lesssim
10^{11}$~G, which is at odd with the hints for anisotropic
distribution of temperature inferred from the analysis of their pulse
profiles.  For example, 1E~1207.4-5209, which displays a single,
rather symmetric pulse profile in the 0.5-2.5~keV band (see Fig.~13 in
\citealt{gotthelf13}), has a PF of about 10\%.  PSR~J1852+0040
\citep{halpern10} also displays a single peak and a very large pulsed
fraction of 64\%, clearly problematic for the CCOs. These puzzling
observations have called into question the presence of a strong
internal toroidal field \citep{halpern10,shabaltas12,vigano12b}.  Our
findings of \S~\ref{sec:spectra} support this suggestion. We have
showed that high PFs, reaching 50-65\% for some particular cases with
strong $B_t$, can be achieved, and correlate with a single thermal
peak, as well as with a smaller inferred BB radius. In particular, we
can consider again as a qualitative representative example (since the
simulations are not tailored to specific objects) model B14t at
50~kyr.  This has a PF of 66\% and a BB radius  $\lesssim 3$~km
for $N_h\gtrsim 5\times 10^{21}$~cm$^{-2}$, as shown in
  Fig.~\ref{fig:fits_nh}.  Note that the presence of an atmospheric
  color correction factor (not included in our modeling) would make
  this emitting region to appear even smaller, hence further
  ameliorating the issue of the small emitting areas seen in some of
  these objects (such as PSR~J1852+0040 indeed, \citealt{halpern10}).

The CCO PSR~J0821-4300 in Puppis~A  also displays a clear single peak in the
thermal component.  However, phase-resolved spectroscopy shows that
the 0.5-1~keV pulsed flux is offset by half a phase cycle with respect
to the 1.5-4.5~keV band \citep{gotthelf09}.  Detailed modeling
by \cite{gotthelf10} showed that two well separated, antipodal
regions of different sizes and temperatures can closely match the
observations. As discussed above, such a strong temperature anisotropy
must imply a strong internal { magnetic} component, which contrasts the
low inferred dipolar field of about $10^{11}$~G (Gotthelf et
al. 2013).  At a qualitative level, we note that a temperature profile
similar to that of model B14t at age $t=5\times 10^4$~yr has some key features that could
mimic what observed for this object. At phase $\gamma=0$, the emission
from this object is in fact dominated by the hot, compact region at
$\theta\lesssim 20^\circ$, and would hence appear as a small hot
spot. On the other hand, when the star is at phase $\gamma=180^\circ$,
only the cooler, more extended region that covers the star for $\theta
\gtrsim 80^\circ$ (cfr. bottom panel of Fig.~\ref{fig:temp2}, dashed line) 
is visible to the observer.  The effective flux
to the observer would mimic that of two antipodal regions, of different
temperatures. We reserve to future work a more comprehensive
exploration of initial magnetic field configurations that can yield
special temperature anisotropy patterns tailored to specific objects.

\section{Summary}\label{sec:conclusion}

We have used results from detailed simulations of the magnetothermal evolution
of isolated NSs to explore the temperature maps, and the resulting spectra and pulse profiles, at
different NS ages, for a variety of magnetic field configurations and
strengths.  In particular, we contrasted models with a purely poloidal
(dipolar) field at birth, with cases in which the initial magnetic
field is dominated by a strong toroidal dipolar component ($\gtrsim
90\%$ of magnetic energy stored in the toroidal field). Intermediate
cases, where $\lesssim 50\%$ of magnetic energy is stored in the
toroidal field, are very similar to the case with an initial purely
poloidal field.

We have found that poloidal-dominated configurations can only yield
symmetric pulse profiles (as expected, for a wide range of viewing
geometries) and very low pulsed fractions unless the local emission is
highly beamed. On the other hand, the anisotropies in the surface
temperature profile induced by the presence of a strong, large-scale toroidal field
generally produce single pulsed profiles, with a pulsed fraction that
can reach the 50-60\% level even for purely isotropic local
emission. These results are particularly relevant to the
interpretation of the thermal spectra of magnetars, which are
often single peaked and highly pulsed. A strong internal component is
in fact believed to play an important role in enhancing their X-ray
luminosities, and in producing a non-negligible outburst rate, even in
'low field' magnetars.  Another class of objects for which our results bear
direct relevance is that of the CCOs. Although their inferred external
(dipolar) fields are very low, some of them present strong hints for a strong surface temperature
anisotropy. Representative cases from our simulations display a
qualitative behaviour consistent with the observational properties of
these objects.

Another finding of our work is that the theoretical thermal X-ray
spectra, when simulated as "real" spectra and then fitted with a
single blackbody (as routinely done in spectral studies) yield
blackbody radii which can be either comparable to, or smaller than,
the actual NS radius, depending on the magnitude of the surface
temperature anisotropy and the absorption column.  Generally, older
objects are more anisotropic, and their inferred BB radii are
smaller. A strong large-scale toroidal field further increases the
surface temperature anisotropy; this, coupled with strong interstellar
absorption, can lead to 'measured' BB radii of $\sim$ 1-2 \~km in
size.  This is again of high relevance for the interpretation of X-ray
observations of isolated NSs, such as e.g.  magnetars, which very
often display blackbody radii of only a few km in size.

As a further related consequence, caution should be exercised when using
the radius derived from spectral fitting, in combination with a mass
or redshift measurement, to constrain the properties of the equation
of state of dense matter. Even for objects with a measured low
external magnetic field, a strong internal component can exist, and hence influence
the delicate measurement of the size of the emitting region.

\section*{Acknowledgements}
This research was supported by 
NSF grant No. AST 1009396 and NASA grants AR1-12003X, DD1-12053X, GO2-13068X, GO2-13076X
(RP), grants AYA 2010-21097-C03-02 (DV, JAP),
AYA2009-07391, AYA2012-39303, SGR2009-811, and iLINK 2011-0303 (NR).
DV is supported by a fellowship from the \textit{Prometeo} program for
research groups of excellence of the Generalitat Valenciana
(Prometeo/2009/103) and NR is supported by a Ramon y Cajal Fellowship.
DV thanks JILA (Boulder, CO, USA) for its kind hospitality during the
time that some of this work was carried out. { Last, we thank an anonymous referee
for very helpful comments.}

\bibliography{ms}

\end{document}